\documentclass[10pt,conference]{IEEEtran}
\usepackage{soul}
\usepackage{amsmath}
\usepackage{amsfonts}
\usepackage{algorithmic}
\usepackage{graphicx}
\usepackage{textcomp}
\usepackage{balance}
\usepackage{fontawesome}
\usepackage{xcolor}
\usepackage{booktabs}
\usepackage{lipsum}
\usepackage{subcaption}
\usepackage{multirow}
\usepackage{wrapfig}
\usepackage{fancyhdr}
\usepackage[many]{tcolorbox} 
\usepackage{pifont}
\usepackage{enumitem}
\newcommand{\proj}{\textsc{FIBER}}
\newcommand{\specialcell}[2][c]{%
            \begin{tabular}[#1]{@{}c@{}}#2\end{tabular}}
\newcommand{\Fig}[1]{Fig.~\ref{#1}}
\newcommand{\Equ}[1]{Equ.~\ref{#1}}
\newcommand{\Tbl}[1]{Tbl.~\ref{#1}}
\newcommand{\Sec}[1]{Sec.~\ref{#1}}

\newcommand{\hpcayear}{2027}

\newcommand{\hpcasubmissionnumber}{139}
\title{A Thread-Register Decoupled GPU Execution Model for Efficient Tensor Computation}

\def\hpcacameraready{} 

\newcommand\hpcaauthors{Zihan Liu$\dagger$, Jingwen Leng$^{*}\dagger$, Yangjie Zhou$\ddagger$, Yitong Ding$\dagger$, Guanlin Zhu$\diamondsuit$, Yilu Huang$\dagger$, Chiheng Jin$\dagger$, \\Chen Zhang$\dagger$, Shixuan Sun$\dagger$, Yu Feng$\dagger$, Anbang Wu$\dagger$, Minyi Guo$\dagger$, Jian Weng$\spadesuit$, Jiajin Tu$\clubsuit$, Junsong Wang$\clubsuit$}
\newcommand\hpcaaffiliation{Shanghai Jiao Tong University$\dagger$, National University of Singapore$\ddagger$, ShanghaiTech University$\diamondsuit$, KAUST$\spadesuit$, HuaWei$\clubsuit$}
\newcommand\hpcaemail{$\lbrace$altair.liu, dingyitong, huangyilu, wendy-hamlet, chenzhang.sjtu, sunshixuan, y-feng$\rbrace$@sjtu.edu.cn, yj\_zhou@nus.edu.sg\\ $\lbrace$ leng-jw$^{\text{\faEnvelope}}$, anbang, guo-my $\rbrace$@cs.sjtu.edu.cn, jian.weng@kaust.edu.sa, tjjsh2109@126.com, junsongwang@huawei.com}

\author{
  \ifdefined\hpcacameraready
    \IEEEauthorblockN{\hpcaauthors{}}
      \IEEEauthorblockA{
        \hpcaaffiliation{} \\
        \hpcaemail{}
      }
  \else
    \IEEEauthorblockN{\normalsize{HPCA \hpcayear{} Submission
      \textbf{\#\hpcasubmissionnumber{}}} \\
      \IEEEauthorblockA{
        Confidential Draft \\
        Do NOT Distribute!!
      }
    }
  \fi 
}

\fancypagestyle{camerareadyfirstpage}{%
  \fancyhead{}
  
  \fancyfoot[C]{}
}
\begin{document}
\maketitle

\ifdefined\hpcacameraready 
  \thispagestyle{camerareadyfirstpage}
  \pagestyle{empty}
\else
  \thispagestyle{plain}
  \pagestyle{plain}
\fi

\newcommand{\hpcaheight}{0mm}
\ifdefined\eaopen
\renewcommand{\hpcaheight}{12mm}
\fi

\begingroup
\renewcommand{\thefootnote}{*}
\footnotetext{Corresponding author.}
\endgroup
\newtcolorbox{takeaway}{
  enhanced,
  sharp corners,
  colback=white,        
  colframe=black!60,    
  boxrule=0.5pt,        
  left=1pt,             
  right=1pt,
  top=1pt,
  bottom=1pt,
  before skip=2pt,      
  after skip=2pt,
  width=\linewidth,     
}

\begin{abstract}
Modern GPUs increasingly integrate Tensor Cores into the execution pipeline. Although aggregate tensor throughput continues to grow, aided by an operand supply that has evolved from register-based in Ampere to redundancy-free, memory-based in Hopper and Blackwell, efficiently orchestrating the complete tensor compute pipeline for the modern AI workloads remains challenging.
We identify the fundamental bottlenecks as fixed parallelism and coarse-grained scheduling, both of which are exposed by modern AI workloads that interleave diverse non-GEMM operations with GEMM. 
To orchestrate tensor computation efficiently, we propose \proj{}, a new architecture that extends the GPU SIMT (single instruction, multiple thread) model. Its basic execution instance, the \emph{fiber}, is decoupled from private register ownership, carrying only minimal control state while accessing an SM's registers through a shared view. 
This enables dynamic parallelism scaling, fine-grained register-level dataflow scheduling, and offers a redundancy-free alternative for matrix operand supply.
We extend the ISA, microarchitecture, and compiler to realize shared-register addressing, conflict-free operand delivery, and fiber-based program mapping. 
Under a typical mixed-precision LLM serving scenario, \proj{} achieves a \textbf{2.25$\times$} end-to-end speedup on Ampere (\textbf{1.15$\times$} for the original FP16 computation), with \textbf{1.8$\times$} and \textbf{2.09$\times$} on Hopper and Blackwell respectively, and kernel-level gains up to \textbf{2.49$\times$}.
\end{abstract}

\section{Introduction}
The virtuous cycle between algorithmic innovation and GPU architecture facilitates the rapid evolution of artificial intelligence (AI) models such as large language models (LLMs).
GPUs introduce new hardware capabilities~\cite{Ampere,Hopper,Blackwell} that AI researchers can exploit to develop new operators~\cite{FlashAttention1,MLA,ClusterFusion,MPK,FlashFuser} and even new model structures~\cite{MOE,NSA}.
These algorithmic improvements, in turn, influence hardware design.
The efficient and programmer-friendly SIMT (single instruction, multiple thread) execution model lies at the core of this virtuous cycle.
The SIMT model exposes massive parallelism, and allows programmers to program from a single-thread perspective.
The hardware groups threads into warps for efficient lockstep execution, and requires a large register file structure to privately hold threads' execution context.

As matrix multiplication (GEMM) increasingly dominates modern AI and LLM workloads, GPUs integrate dedicated matrix accelerators such as NVIDIA's Tensor Cores (TCs) and AMD's Matrix Cores~\cite{Volta,AMD}.
To satisfy the growing TC compute throughput and its corresponding demand for operands, the matrix operand-supply approach has continuously evolved toward a memory-based supply in Hopper and Blackwell, avoiding the warp-level redundant matrix-fragment loading of the private-register supply in Volta and Ampere~\cite{Duplo} (redundancy-free). 
As shared and tensor memory are shared within an SM, operands can be reused across warps (\Sec{sec:moti}).
This lets operand supply keep pace with the rapidly rising compute throughput, achieving near optimal TC utilization.

However, the modern AI workloads are growing more complex.
Attention and its variants~\cite{Attention,KDA,GDN}, mixed-precision kernels~\cite{AWQ,qServe}, feed forward network (FNN) blocks~\cite{GLU} interleave diverse non-GEMM operations with GEMM.
Our analysis shows that although TCs are now well integrated into GPUs, the orchestration of the full tensor computation pipeline for these workloads remains suboptimal, a limitation rooted in the current thread-based SIMT execution model with private register ownership.
This model fixes the number of threads and their register budget at launch time. Yet modern workloads alternate between memory-, tensor-, and vector-bound phases, each demanding a different degree of parallelism and register occupancy, and a static allocation cannot adapt to these phase-dependent needs. The existing CUDA Dynamic Parallelism~\cite{CDN} relies on nested kernel launches with significant runtime overhead, rather than dynamically adjusting thread number and register budget within a kernel.

The model also precludes efficient warp-level dataflow scheduling.
On one hand, SM-scope shared memory provides the semantics for warps within a block to communicate, but it necessitates coarse-grained, phase-based scheduling with no fine-grained dependency tracking. On the other hand, fixed-latency compute instructions with their stall bits, and per-warp dependency counters can support fine-grained dependency tracking~\cite{Dissecting}, but these mechanisms are confined to each warp's private registers, restricting fine-grained dataflow to within a single warp.
Ideally, these workloads require dynamic parallelism that adapts across execution phases, fine-grained dataflow scheduling, and preserving a redundancy-free TC operand supply. Meeting these goals requires rethinking the binding between execution instances and registers.

We propose \proj{}, an architecture built on a decoupled, shared register execution model that extends SIMT while remaining fully compatible with it.
\proj{} introduces a new parallel execution instance, the \textbf{fiber}, analogous to a SIMT thread but without private register ownership, carrying only minimal state such as the program counter and status flags.
Unlike SIMT threads, which own their registers privately, fibers access an SM's register file through a shared view.
This design lets the hardware scheduler adjust the number of active fibers at runtime, no longer bounded by a pre-determined private register allocation, supporting larger resident parallelism with tiny contexts and matching the parallelism demands of different compute phases.
The shared register view also enables direct register-level dataflow and fine-grained scheduling, backed by a per-register dependency-tracking mechanism.
Shared registers can further serve as a redundancy-free operand-supply path for matrix computation to guarantee the optimal GEMM performance.
Together, these features let \proj{} orchestrate tensor computation efficiently.

We design a \proj{}-based GPU with minimal modifications and largely reuse existing hardware architectures.
First, we extend the current GPU ISA to support shared-register semantics, dynamic parallelism, and fine-grained scheduling.
Second, we introduce lightweight microarchitectural enhancements that preserve the existing physical register organization, including a selector and a crossbar for shared register access, an arbiter to resolve the port contention caused by shared access, and a register busy bitmap for fine-grained cross-fiber dependency tracking.
Third, we present the \proj{} programming model which exposes a shared-register view, enabling fibers to communicate through registers and orchestrate fine-grained register dataflow. It also remains SIMT-compatible.

We evaluate \proj{} on LLM workloads using GTSim~\cite{GTSim}, a tile-graph based cycle-level simulator.
\proj{} delivers \textbf{2.25$\times$}, \textbf{1.8$\times$} and \textbf{2.09$\times$} end-to-end speedup under typical mixed-precision LLM serving scenarios against Ampere, Hopper and Blackwell baselines. 
With kernel level speedup up to \textbf{2.49$\times$}.
The area and power overhead are less than 0.1\% and 0.4\%.
A further overhead analysis shows that, even under an extremely tight hardware budget, the key enabler of \proj{}'s speedups remains feasible.
These results show that \proj{} provides efficient orchestration for tensor computation.

We summarize our contributions as follows:
\begin{itemize}[leftmargin=*]
\item A comprehensive analysis of tensor computation bottlenecks on GPUs: static parallelism and coarse-grained scheduling.
\item The design of \proj{}, an architecture that decouples register ownership from parallel execution instances, comprising ISA extensions, lightweight microarchitectural enhancements, and a CUDA-compatible programming model.
\item 1.8-2.3$\times$ end-to-end speedup over the multiple GPU baselines under typical LLM serving scenarios, preserving TC performance while closing the gap to theoretical peak.
\item A thorough ablation study quantifying the speedup contributed by each of \proj{}'s key features.
\end{itemize}
\section{Background}
\label{sec:background}
In this section, we first present background on LLM workloads and their growing complexity. Next, we explain the basic GPU hardware architecture and its programming method. 

\subsection{LLM Workloads}
Large language models (LLMs) mostly adopt transformer-based architectures. The core component of a transformer block is the attention mechanism~\cite{Attention}, whose basic multi-head formulation is given in \Equ{equ:attention}. Attention is one of the most heavily optimized kernels, as exemplified by the FlashAttention series~\cite{FlashAttention1,FlashAttention2,FlashAttention3}, which leverages GPU features to perform efficient fused computation.
\begin{equation}
\label{equ:attention}
\begin{aligned}
\emph{Attention}(\emph{Q}, \emph{K}, \emph{V})
= \emph{Softmax}\left(\emph{Q}\cdot \emph{K}^{\top}/\sqrt{\emph{d}}\right)\cdot \emph{V}
\end{aligned}
\end{equation}

Before stateful attention, a linear projection transforms the hidden states from the previous layer into query ($Q$), key ($K$), and value ($V$) states, each produced by a GEMM with the corresponding weight matrix.
After attention, the resulting context vector is passed through an output projection, again implemented as a GEMM.
Then the activations flow into a feed-forward network (FFN), which is dominated by GEMM layers interleaved with lightweight nonlinear activations~\cite{GLU}; the FFN output then becomes the input to the next transformer layer.
Although additional components such as RMSNorm~\cite{RMSNorm}, RoPE~\cite{RoPE}, and Alibi~\cite{Alibi} appear in many architectures, GEMM operations remain the overwhelming computational bottleneck in modern LLMs.

To cope with rapidly increasing model sizes, modern LLMs employ a range of optimizations, including efficient attention and \emph{mixed-precision computation}, commonly denoted as $\text{W}x\text{A}y$ to indicate $x$-bit weights and $y$-bit activations.
In this work, we consider both the typical compute phases of LLM architectures and these widely adopted optimizations.

\subsection{Basic GPU Architecture and Programming}
GPUs employ massive thread-level parallelism to deliver high computational throughput.
A GPU comprises multiple processing blocks, known as Streaming Multiprocessors (SMs), which communicate through off-chip global memory.
In this work, we use NVIDIA GPUs as the baseline.
Since Volta, each SM is organized into four subpartitions (SP) that share a unified L1 cache and shared-memory space~\cite{Volta,Ampere,Hopper,Blackwell}.
Within each subpartition, instructions are first handled by a scheduler and dispatcher, then issued to one of several pipelines, including the CUDA cores (SIMD units), Tensor Cores, Special Function Units (SFUs), and Load/Store units.
To program an NVIDIA GPU, developers follow the SIMT (single instruction, multiple thread) model in which all threads conceptually execute the same instruction sequence.
A program is decomposed into independent \emph{thread blocks}, each of which is scheduled onto an SM; multiple blocks may run concurrently on the same SM to hide latency.
Threads within a block are grouped into \emph{warps} of 32 threads (the basic scheduling unit), which execute in lockstep.
\begin{figure*}[t]
    \centering
    \includegraphics[width=0.99\linewidth]{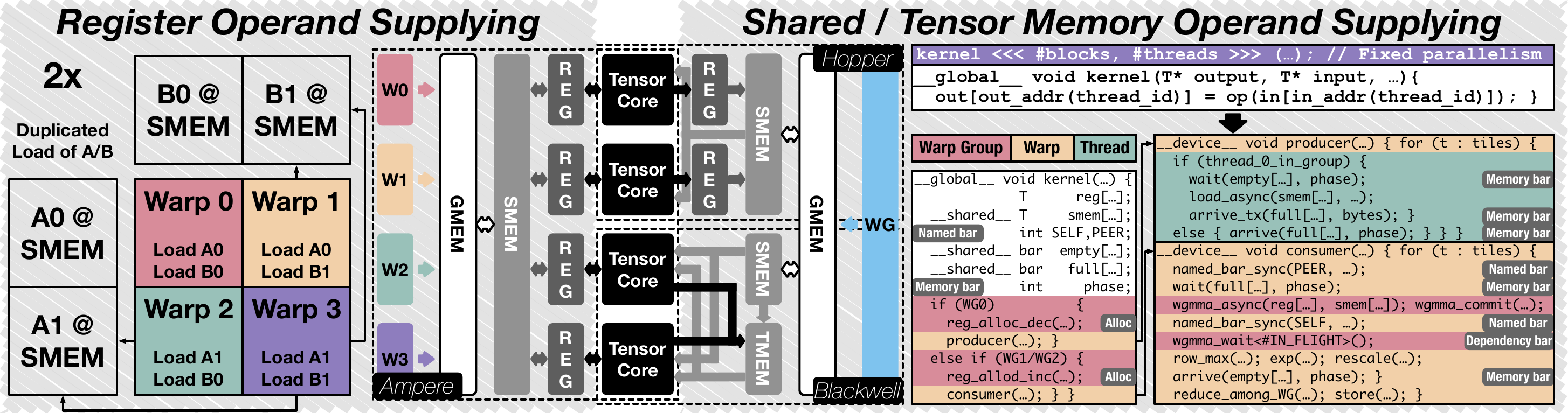}
    \vspace{-0.2cm}
    \caption{Two methods of TC operand supply. (left) register operand supply. (right) shared/tensor memory operand supply.}
    \vspace{-0.2cm}
    \label{fig:integration}
\end{figure*}
\section{Challenge of Efficient Tensor Computation}
\label{sec:moti}
Modern GPUs increasingly integrate Tensor Cores (\textbf{TC} hereon) or Matrix Cores for high matrix-computation throughput~\cite{Volta,Ampere,Hopper,Blackwell,MI300}.
To meet the rising operand supply demands of this growing throughput, NVIDIA converged on a shared/tensor-memory operand supply starting with the Hopper generation, eliminating the duplicated matrix-operand loads of the register-supply approach used in Volta and Ampere~\cite{SHREG,LARF}. As \Fig{fig:integration} shows, under register supply warp 3 reloads fragments $A_1,B_1$ already loaded by warps 1 and 2, since warps cannot exchange data through registers; removing this duplication lifts a key obstacle to TC scaling.
Memory supply enables optimal TC throughput on GEMM. However, the modern workloads, especially LLMs, are not pure GEMM: diverse non-GEMM operations on the SIMD/SFU pipelines are interleaved with GEMM, as in attention, FFN, and more complex variants. Efficient TC integration alone is therefore not enough; the GEMM must also be orchestrated efficiently alongside other non-GEMM operations. In this section, we identify two main bottlenecks in this orchestration.

\subsection{Static Parallelism}
\begin{figure}[t]
    \centering
    \includegraphics[width=0.32\linewidth]{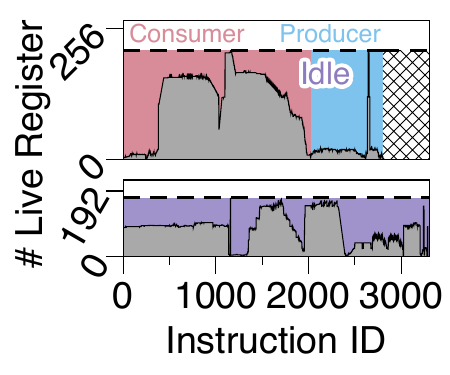}
    \includegraphics[width=0.32\linewidth]{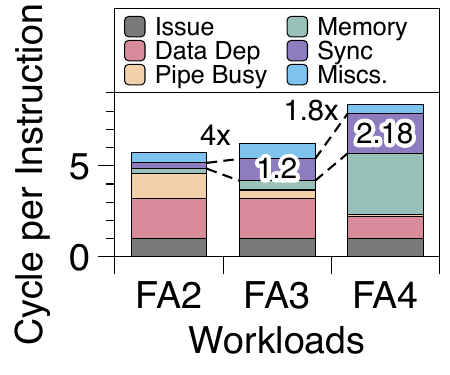}
    \includegraphics[width=0.32\linewidth]{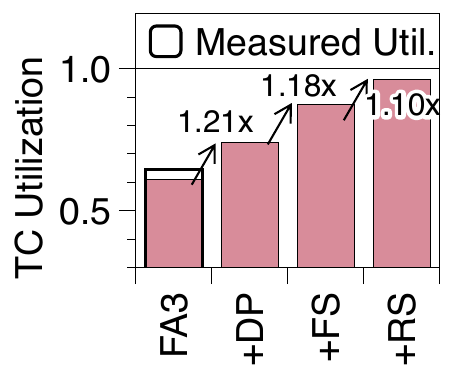}
    \vspace{-0.2cm}
    \caption{(a) Live register trace of FA3 (upper) and FA4 (lower). (b) Cycle-per-instruction breakdown of FA2/3/4. (c) Concept verification experiment of Attention on simulated H100.}
    \vspace{-0.2cm}
    \label{fig:motivation}
\end{figure}

Both operand supply approaches preserve the single instruction, multiple thread (SIMT) execution model: each thread privately owns its registers, and both the per-thread register allocation and the number of active threads are fixed at launch. CUDA Dynamic Parallelism~\cite{CDN} can spawn additional work at runtime, but only through nested kernel launches with substantial launch and runtime overhead.
For pure GEMM, this is not a problem: it is not parallelism-hungry, so a good-enough operand supply can provide optimal throughput. While for complex workloads like attention, GEMM is interleaved with many non-GEMM operations that execute on the SIMD or SFU pipelines. GEMM is not parallelism-hungry but has high operand-supplying demands, whereas the other operations leave much of the register budget idle yet are starved of thread-level parallelism, leaving the pipeline underutilized.
Although the Hopper generation introduces warp specialization and a register re-allocation mechanism (\texttt{setmaxnreg}), such idleness and waste persist, as shown in the upper panel of \Fig{fig:motivation}(a) (FA3): even though the producer and consumer are allocated different register budgets, the consumer still suffers register idleness and insufficient parallelism during the softmax phase. Within the consumer group, the most register-hungry instructions, which occupy more than 200 registers, account for only 16.5\% of all executed instructions.
Even on Blackwell, which bypasses registers entirely for its TC computation, the static parallelism still cannot satisfy the different compute stages surrounding the TC (lower panel of \Fig{fig:motivation}(b), FA4).
\begin{takeaway}
\textbf{Takeaway 1:} The private register ownership of the SIMT model limits parallelism scaling across compute phases.
\end{takeaway}

\subsection{Coarse-grained Scheduling}
With the added TC compute pipeline, coordinating dataflow across pipelines is critical to high-performance kernels~\cite{Tawa,Kitsune}. 
Since Hopper, TC-related dataflow has been organized at warp-group granularity through phase-based, coarse-grained bulked dependency tracking. This differs from the earlier warp-granularity scheduling, which relied on the fixed-latency guarantees of compute instructions, stall bits, and per-warp dependency counters, but at the cost of operand redundancy and the absence of cross-warp reuse.
This limits fine-grained data reuse and forwarding, and forces programmers to reason about additional coarse-grained abstraction layers. Beyond SIMT-aligned warp abstraction and its \texttt{cp.async.commit/wait} dependency barrier, programmers must now reason about warp-group (four-warp) coordination and, on Blackwell, thread-block clusters for two-SM operand sharing. Synchronization grows correspondingly complex, additionally requiring memory barriers between warp-specialized producer and consumer groups (\texttt{mbarrier.arrive/wait}) and named barriers within consumer warp groups (\texttt{bar.sync~\%id}). These abstractions and synchronization mechanisms not only increase programming complexity, as reflected in the lines of code in \Fig{fig:integration}, but also introduce increased synchronization stalls and barrier overhead. As \Fig{fig:motivation}(b) shows, synchronization stalls account for a rapidly growing fraction of total CPI (26\% for FA4). This cost also surfaces in the originally reported numbers: although FA3 achieves higher absolute performance~\cite{FlashAttention3}, its TC utilization is lower than FA2's~\cite{FlashAttention2}.
\begin{takeaway}
\textbf{Takeaway 2:} Coarse-grained scheduling introduces overhead in both programming and execution.
\end{takeaway}

\subsection{Optimal Dataflow Analysis}

Based on the previous discussion, an ideal tensor-computation dataflow should support: (i) dynamic parallelism that adapts the number of parallel executing instances to different computation phases, and (ii) fine-grained scheduling with minimal synchronization overhead. It should also preserve redundancy-free operand supplying for TCs. To quantify these benefits, we conduct an oracle analysis of FA3 using GTSim configured to model an NVIDIA H100 GPU~\cite{GTSim}.

\textbf{Dynamic Parallelism (DP).}{ }
Although FA3 employs warp specialization and dynamic register reallocation, the thread count of each warp-group remains fixed across computation phases. To evaluate dynamic parallelism, we assume an unbounded physical register file and allow the scheduler, when the head instruction is stalled, to issue the earliest ready instruction targeting an otherwise idle pipeline, thereby approximating oracle-level dynamic parallelism. \textbf{DP} enables SIMD and SFU operations within the consumer group to exploit greater thread-level parallelism, yielding a 1.21$\times$ speedup.

\textbf{Fine-Grained Scheduling (FS).}{ }
To evaluate FS, we leverage GTSim's tile-graph representation to decompose each warp-group-level TC node into four warp-level TC nodes, each operating on one quarter of the original tile while retaining shared-memory operand supplying. The baseline warp-group-wide, phase-level bulk synchronization is replaced with dependency edges between warp-level computation nodes and a new per-warp data tile scoreboard, enabling fine-grained dependency tracking. This flexibility yields a 1.18$\times$ speedup.

These features naturally motivate a register-oriented sharing design. First, \textbf{DP} favors a loosely coupled register-thread paradigm, as relaxing private register ownership decouples threads from static register budgets. Second, the physical register file is already organized in a 32-lane, warp-aligned manner, matching the desired dependency-tracking granularity of \textbf{FS}. As an additional benefit, register sharing (\textbf{RS}) preserves redundancy-free operand supplying and enables direct register dataflow. We therefore extend register addressing across subpartitions and reroute operand supplying and reduction dataflow to the shared register, yielding a 1.1$\times$ speedup.

\begin{table}[b]
    \caption{Comparing \proj{} with existing generations}
    \vspace{-0.2cm}
    \label{tab:feature}
    \centering
    \footnotesize
    \begin{tabular}{cccc}
    \toprule
    \textbf{Feature} & \specialcell{Pre-\\Hopper} & \specialcell{Hopper \&\\Blackwell} & \textbf{\proj{}} \\
    \midrule
    Parallelism (Single Kernel) & Static & Static & \textbf{Dynamic} \\
    Orchestration Granularity & Warp & Warp-group & \textbf{Weft (Warp)} \\
    Register Ownership & Private & Re-allocation & \textbf{Shared} \\
    Cross-warp Dataflow & Mem & Mem & \textbf{Reg, Mem} \\
    TC operands supplying & Reg & Reg, Mem & \textbf{Reg, Mem} \\
    \bottomrule
    \end{tabular}
\end{table}

\proj{} adopts a shared-register design by default, as it provides the cleanest abstraction. Under an extremely tight hardware budget for wiring or crossbar, \proj{} guarantees \textbf{DP} and \textbf{FS}, and a limited, intra-subpartition register sharing (detail analysis in \Sec{sec:hwoverhead}). \Tbl{tab:feature} summarizes \proj{}'s key features against existing GPU generations.

\section{\proj{} Overview}
\label{sec:overview}
We propose \proj{}, an architecture built around lightweight parallel execution instance called \textbf{fiber} that delivers desirable features identified above for efficient tensor computation. We present an overview of the execution model, ISA extensions, microarchitecture, and software stack.

\begin{figure}[t]
    \centering
    \includegraphics[width=0.99\linewidth]{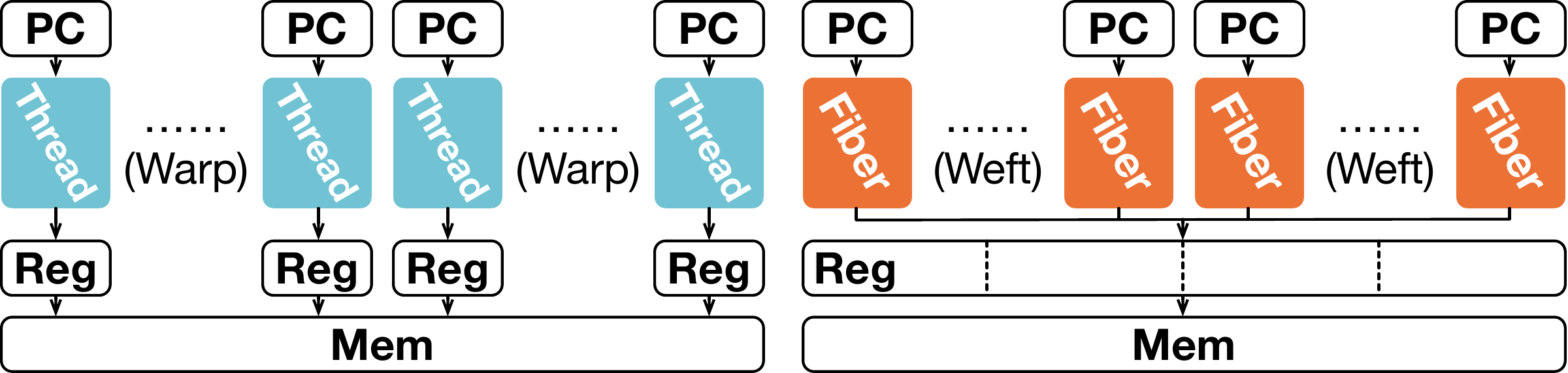}
    \vspace{-0.6cm}
    \caption{Comparison between SIMT-thread and \proj{}-fiber.}
    \vspace{-0.2cm}
    \label{fig:overview}
\end{figure}
\textbf{\proj{} Execution Model.}
A \textbf{fiber} is the basic parallel execution instance, analogous to a SIMT thread but carrying minimal architectural state (a program counter, identifier, and control flags) \ul{\emph{\textbf{without owning private registers.}}} Instead, all fibers resident on the same SM access the register file as a unified, shared resource. \Fig{fig:overview} contrasts a conventional thread with a fiber. 
By decoupling register ownership from execution, this design breaks the tight coupling between register budget and the number of parallel execution instances (fibers): schedulers can scale the number of active fibers up or down while maintaining only a small scheduling context, exposing dynamic-parallelism semantics.
Fibers that execute the same instruction in lockstep form a \textbf{weft}, analogous to a SIMT warp but with shared-register semantics. This shared substrate supports fine-grained register-level dataflow scheduling, letting fibers cooperate directly so that data is forward and reused in registers rather than shuttled repeatedly across the memory hierarchy. Such shared registers can also serve as a redundancy-free operand supply source for the TCs.

\textbf{ISA Extension.}
We introduce a minimal ISA extension to support register addressing under the fiber's decoupled register model (\Fig{fig-isaextension}).
We retain the contiguous 32-lane vector register access pattern to align with existing register organization and compute/memory pipelines.
With decoupled ownership, an instruction can address physical registers of all subpartitions within an SM.
Each instruction's register operands are extended with additional bits for cross-subpartition addressing, enabling shared-register access as a natural consequence of the decoupled model.
An additional instruction is added to support dynamic scaling of fiber count, and an existing instruction is extended to enable fine-grained register-level dataflow.

\textbf{Microarchitecture.}
At the microarchitectural level, \proj{} preserves the existing GPU register file organization, avoiding costly redesign.
The restricted register access mode described above (32-lane aligned) maintains compatibility with current bank and crossbar structures.
Lightweight hardware units are introduced to enable shared register access across an SM, together with an arbitration unit to handle the new cross-subpartition port conflicts.
A register-grained cross-fiber dependency tracking unit is also introduced.
All mechanisms require minor modifications to the baseline design.
As a result, \proj{} achieves higher register utilization and improved TC throughput with minimal hardware overhead, providing a practical path toward fiber-enabled GPU architectures.

\textbf{\proj{} Software Stack.}
Building on the decoupled register abstraction, a \emph{\proj{}-native} programming mode is provided, which exposes a shared, unified register view within an SM. This view enables dynamic parallelism and fine-grained register-level dataflow through a few \proj{}-specific primitives. Several compiler passes are introduced to coordinate the shared register access. 
Moreover, with the address extension ignored, a fiber behaves exactly like an SIMT-thread, accessing its own private register. \proj{} is therefore naturally compatible with existing CUDA programs (\emph{CUDA-native} mode).

\section{\proj{} ISA Extension}
We introduce a set of ISA extensions to support decoupled, shared register model, dynamic parallelism, and fine-grained scheduling, while preserving compatibility with existing SIMT instructions.
We first describe the addressing extension for shared register semantics, then the instructions that enable dynamic fiber count scaling and fine-grained scheduling with register-level dataflow.
Before detailing each component, we present a key architectural insight that guides these extensions.

\subsection{Base ISA Design and \proj{}'s Design Choice}

\begin{figure}[t]
    \centering
    \includegraphics[width=0.99\linewidth]{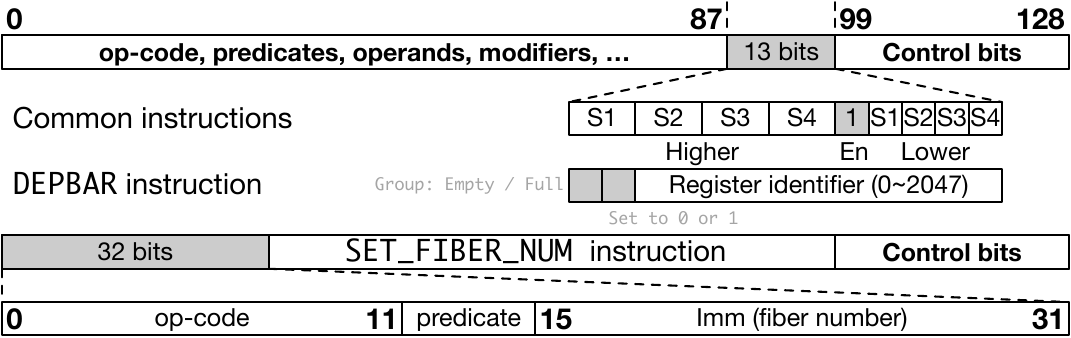}
    \vspace{-0.4cm}
    \caption{Format of extended instructions in \proj{}.}
    \vspace{-0.2cm}
    \label{fig-isaextension}
\end{figure}

Physically, each SM contains 65536~32-bit registers (256~KB in total), organized as 2048~\emph{32-lane vector registers}, aligning with the warp abstraction.
Each subpartition owns 512 of these vector registers, yet SASS~\cite{SASS}, the base CUDA assembly instruction format, can address only 256 of them through 8-bit register operand fields.
Supporting \emph{fully} shared register access across fibers therefore requires extending every register operand to 16 bits.
Because the longest SASS instruction carries one destination and four source operands~\cite{Kuter}, this extension would add 40 bits per instruction.
However, reverse engineering reveals only 13 unused bits in the SASS encoding (top of \Fig{fig-isaextension})~\cite{DecCUDABin,DissectingVolta}, making full-range shared-register addressing infeasible within the existing encoding space.

We make two key observations to enable a practical register-sharing extension within the 13 unused bits of the existing SASS format.  
First, \textbf{fine-grained register addressing is unnecessary}. 
Most instructions naturally access registers in units of 32-lanes, reflecting the hardware's 32-lane vector register organization used by hardware pipelines including TCs.  
Because intra-weft cross-fiber communication can be efficiently handled by existing shuffle networks, we retain this 32-lane alignment.
As a result, we only need 11 bits to encode the 32-aligned registers for each operand in the instruction, with remaining 5 bits automatically zero padded.
Second, \textbf{cross-subpartition register writes are not required in hardware}.  
In the current subpartition-based microarchitecture, allowing a fiber to write directly into a remote subpartition's register file would significantly complicate dependency tracking and write-port arbitration.  
Instead, we disallow remote register writes at the ISA level and rely on compiler (\Sec{subsec:compiler_support}) to transform this behavior to dual remote reads and local writes.  
This preserves the destination operand's existing addressing semantics with no additional bits for destination operands.

\subsection{Supporting Shared Register Semantic}
First we need to support the extended addressing of shared register. 
As discussed earlier, we extend only the \emph{source-operand} addressing for instructions that read from the register file.  
Instructions that operate on special-purpose registers (e.g., extended \texttt{DEPBAR} targeting new dependency tracking mecnahism) are treated separately.  
In the SASS encoding, the reserved region begins at bit~87, aligned to a byte boundary.  
We use the first byte of this region to encode the high-order addressing information for the four source operands.  
Specifically, each source operand receives two additional high bits, which are used to select the correct subpartition.  
An additional bit acts as the enable signal for this selector, indicating whether the operand should access a non-local subpartition.  
The remaining four bits are concatenated with the original source address fields respectively, to index registers within the selected subpartition.  
Together, these extensions provide full access to the 32-aligned 65536 registers while staying within the 13 available bits of the SASS format.

\subsection{Supporting Dynamic Parallelism (DP)}
As fibers no longer own private registers, the number of active fibers/wefts is not constrained by static register allocation.  
This enables a new form of dynamic parallelism where a kernel can increase the fiber count during memory-intensive phases to hide latency, or reduce it during compute-intensive phases to ensure sufficient register capacity per fiber.  
To support this capability, the ISA needs to provide a new instruction for software to communicate the desired parallelism to the scheduler that we describe below.

\textbf{\texttt{SET\_FIBER\_NUM}.}  
As shown at the bottom of \Fig{fig-isaextension}, we assign a unique 12-bit opcode to this instruction.  
The following four bits serve as a predicate field, consistent with the existing SASS format.  
All fibers within an SM must participate in the execution of this instruction, equivalent to a synchronization.  
The next 16 bits encode an immediate value specifying the desired number of fibers.  
After the instruction completes, all subsequent instructions are issued according to the newly configured fiber count.  

\begin{figure}[t]
    \centering
    \includegraphics[width=0.99\linewidth]{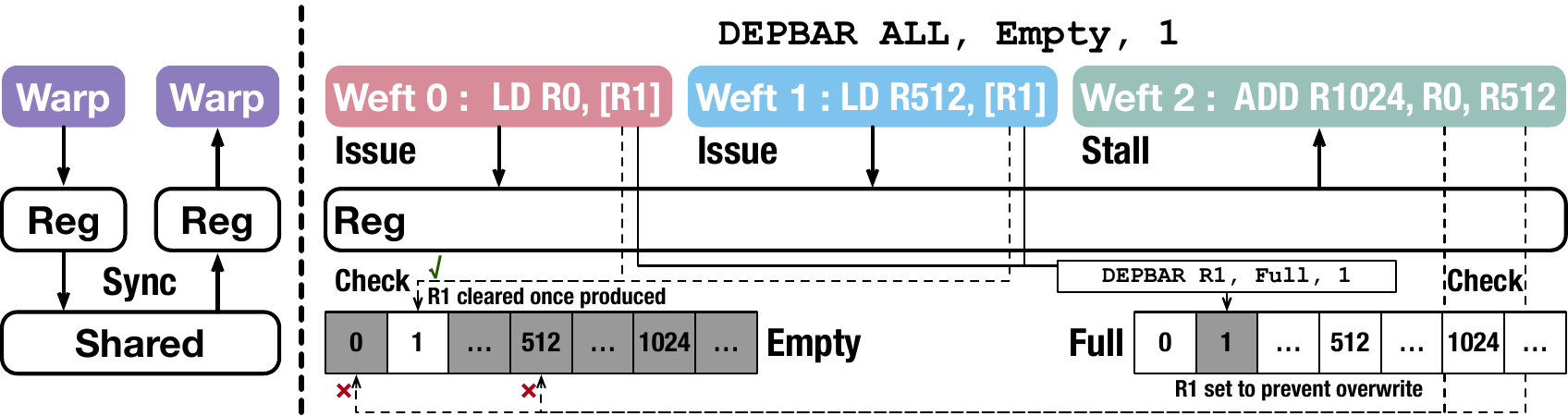}
    \vspace{-0.4cm}
    \caption{Example of cross-fiber/weft dependency tracking.}
    \vspace{-0.2cm}
    \label{fig-depexample}
\end{figure}
\subsection{Supporting Fine-grained Scheduling (FS)}
\proj{}'s shared register semantics lets data flow and be reused across fibers directly through the register file, rather than being staged through memory (\Fig{fig-depexample}, left). 
But shared registers only make such register-level dataflow \emph{possible}; realizing it in a fine-grained manner, without coarse-grained synchronization, requires corresponding fine-grained dependency tracking mechanism.
\proj{}'s hardware provides this tracking with a default, automatic update behavior that enforces the inherent data dependencies in the absence of explicit control, as detailed in \Sec{subsec:FSsupport}; here we focus on the ISA extension that exposes this fine-grained coordination to programmers.

\textbf{\texttt{DEPBAR}. }
Existing GPUs supports \texttt{DEPBAR} instruction that enforces intra-warp dependencies by tracking outstanding operations through dependency slots.
Each warp issues long-latency or asynchronous instructions (e.g., \texttt{cp.async,wgmma.async}) that occupy dependency slots, and \texttt{DEPBAR} stalls execution until the specified slots are resolved.
This design enables fine-grained ordering within a warp while preserving pipeline parallelism~\cite{Dissecting}.

\proj{} extends this instruction to track cross-fiber (weft) dependencies at register granularity, with two states following Cray MTA's terminology~\cite{CrayMTA}: \textbf{\emph{empty}}, where the latest data is not yet ready (empty), and \textbf{\emph{full}}, where the latest data is ready (full) and should be overwritten.
As \Fig{fig-depexample} (right) shows, since there is no ordering guarantee between wefts, all registers are initially marked as  \textbf{\emph{empty}} (not-ready) to prevent reading stale values. Once \texttt{R1} (the source of weft~0/1) is produced and the wefts are issued, they mark \texttt{R1} as \textbf{\emph{full}} (in active use). Meanwhile, weft~2 stalls because its source operands have not been produced (its destination is not in use, so the stall is dictated solely by the source dependency).
Therefore, the ISA must expose, for each shared register, both dependency states along with operations to set and release them.
Our restricted access mode (32-lane alignment) reduces the namespace to 2048 vector registers.
We repurpose the reserved encoding space of \texttt{DEPBAR} as follows: 11 bits encode the identifier of the target vector register, and the remaining two bits specify the group (\textbf{\emph{empty}} / \textbf{\emph{full}}) and the operation (set / clear).

\begin{figure*}[t]
    \centering
    \includegraphics[width=0.99\linewidth]{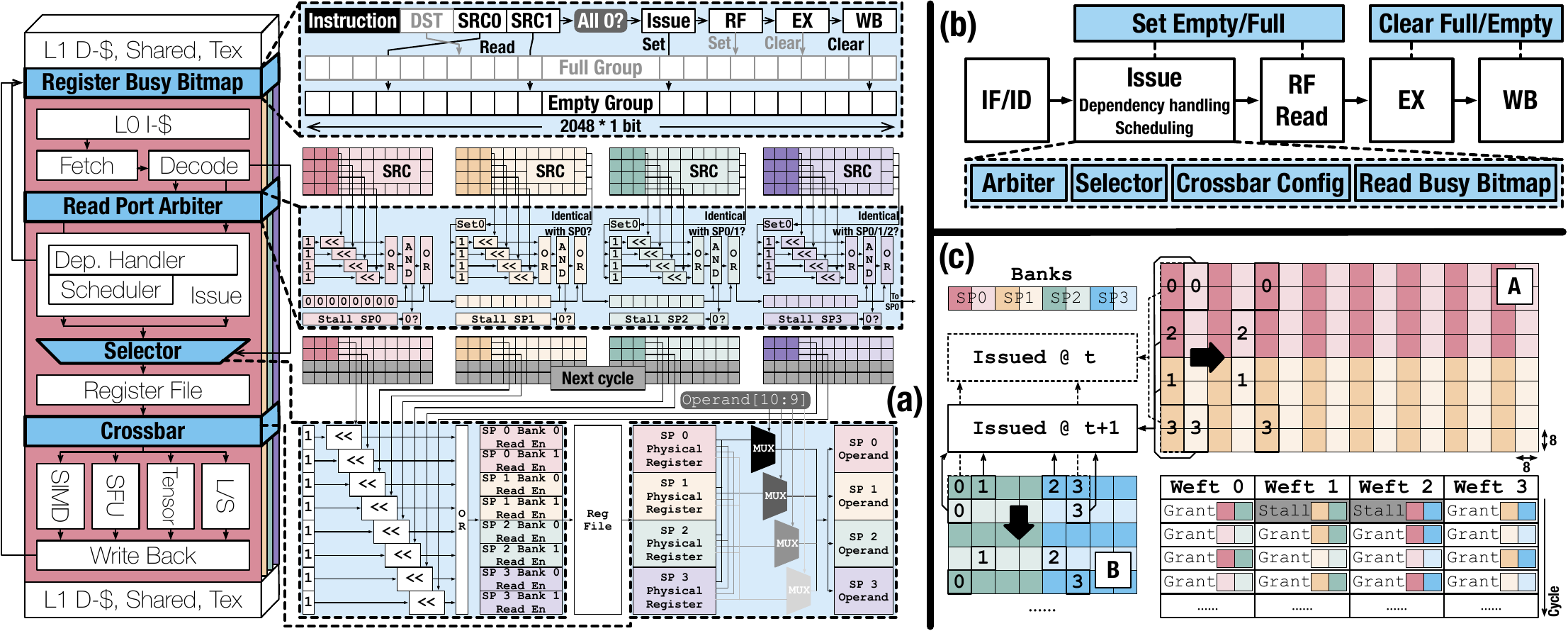}
    \vspace{-0.2cm}
    \caption{(a) \proj{}'s architecture, and micro-architecture of new components. (b) \proj{}'s instruction execution pipeline. (c) Shared register bank conflict and its arbition. }
    \vspace{-0.2cm}
    \label{fig-uarch}
\end{figure*}
\section{\proj{} Micro-Architecture}
To support the \proj{} execution model and corresponding extended ISA, the baseline GPU microarchitecture requires several enhancements.
We first overview the baseline GPU microarchitecture, then detail the enhancements that enable key features of \proj{}.

\subsection{Baseline and Modification Overview}
White boxes in \Fig{fig-uarch}(a) and (b) represent the baseline GPU micro-architecture and instruction pipeline. 
After the IF/ID stage, each instruction resolves its dependencies through a combination of stall bits and hardware dependency counters within the issue logic~\cite{Dissecting}.
Once all dependencies are satisfied, the instruction is dispatched to the register file (RF) read stage.
The instruction then reads its operands, proceeds to execution (EX), and finally writes the results back (WB).

Blue boxes in \Fig{fig-uarch} (a) and (b) represent the necessary extension of \proj{}.
Three main mechanisms are required to support the \proj{} execution model on top of the baseline.
First, the baseline GPU adopts a subpartition-based design where each subpartition accesses only its local physical register file. \proj{} must support access to remote subpartitions' register files. Moreover, shared register access introduces cross-subpartition read-port conflict; we introduce a hardware mechanism to resolve this and preserve the fixed-latency guarantee of compute instructions.
Second, the architecture must support dynamic scaling of fiber count.
Third, it must manage new cross-fiber data dependencies at shared registers level to enable fine-grained register level dataflow scheduling.

\subsection{Supporting Shared Register Access}
\label{sunsec:RS}
We follow the static warp-to-subpartition mapping as baseline GPU does~\cite{Dissecting,Mitigating} to determine the correspondence between wefts and physical register files.
Register sharing necessites remote register access support on the hardware, we introduce a selector module, a crossbar module, and extend the decode logic accordingly.
The selector within each subpartition takes three extended bits of each source operand to determine which of the eight register bank read ports should be enabled. This produces an 8-bit enable vector specifying which banks to access, as illustrated in the bottom of \Fig{fig-uarch} (a).
Once the selected banks return operands, a small 4$\times$4 crossbar per 32 lanes routes the operands from the source subpartition to the requesting subpartition based on the top-2 address bits, also shown in the bottom of \Fig{fig-uarch} (a). The crossbar consists of simple multiplexers and is configured at issue stage.

\textbf{Read Port Conflict Handling. }
Sharing registers introduces a new \emph{inter-subpartition read-port conflict} (every fiber writes only to its local subpartition and write ports never conflict). The SM exposes eight read ports (4 subpartitions $\times$ 2 banks), so four instructions issued together may contend for the same port, producing variable-latency fetches that break the fixed-latency guarantee many instructions require.

We introduce a \emph{Read Port Arbiter} (\Fig{fig-uarch}(a), middle) that generates four stall bits, one per subpartition and merged with the baseline scheduler's decision. Each cycle, it clears an 8-bit \emph{port-occupied register} and scans the subpartitions from a rotating start~$i$ for fairness. Subpartition~$i$ is always granted, with its read ports OR-ed into the register; each subsequent subpartition's ports are AND-ed against it, a non-zero result signals a conflict and stalls it, otherwise it is granted and OR-ed in. Identical-address detection clears the matching occupancy bits when a later operand duplicates one already granted, avoiding needless stalls.
\Fig{fig-uarch}(c) illustrates the arbitration for \texttt{HMMA.FP16.1688}. Since shared-register accesses occur mainly during matrix multiplication, the compiler maps operands by access pattern: Matrix~A spans subpartitions~0--1 row-wise and even/odd banks column-wise, while Matrix~B spans subpartitions~2--3 column-wise and banks row-wise, so each weft reads its $16\times8$ A and $8\times8$ B fragments from single banks. At cycle~$t$, weft~0 is granted and weft~3 does not conflict, while wefts~1 and~2 stall on weft~0's bank. Rotating the start to weft~1 at cycle~$t{+}1$ lets them proceed, and wefts~0 and~3 have meanwhile advanced to new banks, so all four are granted and stay staggered thereafter. The conflict thus self-corrects, costing only one stall per outer iteration. 

\subsection{Supporting Dynamic Parallelism (DP)}
The decoupled nature of fibers enables dynamic parallelism within an SM. The scheduler maintains multiple context entries and can accommodate up to 64 wefts per subpartition, analogous to how a baseline GPU hosts up to 48 warps per SM~\cite{Dissecting}.
Unlike warps, the nature of decoupled register ownership of weft makes large-scale dynamic scaling both practical and lightweight. The storage overhead of these weft contexts (PC, active mask, flags, and related metadata) is at the same magnitude as the baseline GPU.

When the issue logic encounters a \texttt{SET\_FIBER\_NUM} instruction, it first waits for all fibers to reach this instruction as an explicit synchronization point, then reconfigures the number of active wefts to the specified value.
If the current number of active wefts exceeds the requested value, the surplus weft contexts are masked and disabled, meaning they no longer participate in scheduling or issue from the current cycle.
Conversely, if fewer wefts are active than requested, the hardware allocates additional weft contexts, initializes their architectural state, and sets their program counters to the current instruction—this is what we refer to as ``spawning fibers''.
These newly created wefts become eligible for scheduling, enabling the architecture to scale parallelism up or down dynamically to match the demands of compute phases.

\subsection{Supporting Fine-grained Scheduling (FS)}
\label{subsec:FSsupport}
Building on the two states of cross-fiber register dependencies, we introduce an SM-scope \emph{Register Busy Bitmap} for fine-grained register-level dataflow. As shown at the top of \Fig{fig-uarch} (a), it consists of two groups, \textbf{\emph{empty}} and \textbf{\emph{full}}, each with 2048 1-bit entries corresponding to the 2048 physical vector registers, tracking pending writes and in-flight reads respectively.
The bitmap can be manipulated through the extended \texttt{DEPBAR} instruction to coordinate fine-grained register-level dataflow; otherwise, it is maintained automatically by hardware.

An instruction is eligible to issue only when its destination register is clear in the \textbf{\emph{full}} group, preventing overwrites of values still being read by other wefts, and all source registers are clear in the \textbf{\emph{empty}} group, indicating operands are ready. By default, the destination entry is set in the \textbf{\emph{empty}} group at issue to indicate an unready result and cleared at writeback, while source entries are set in the \textbf{\emph{full}} group at RF read to indicate in-use data and cleared after pipeline consumption. Extended \texttt{DEPBAR} instructions can override this default behavior, as detailed in \Sec{sec:programmingmodel}.
The baseline GPU maintains six 6-bit dependency counters per warp~\cite{Dissecting}, and our bitmap maintains a comparable amount of state.

\begin{figure}[h]
    \centering
    \includegraphics[width=0.99\linewidth]{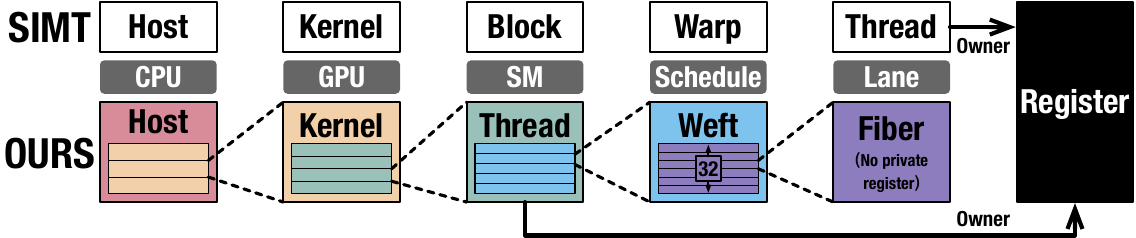}
    \vspace{-0.6cm}
    \caption{The kernel-thread-fiber programming model of \proj{}.}
    \vspace{-0.2cm}
    \label{fig:executionmodel}
\end{figure}

\begin{figure*}[t]
    \centering
    \includegraphics[width=0.99\linewidth]{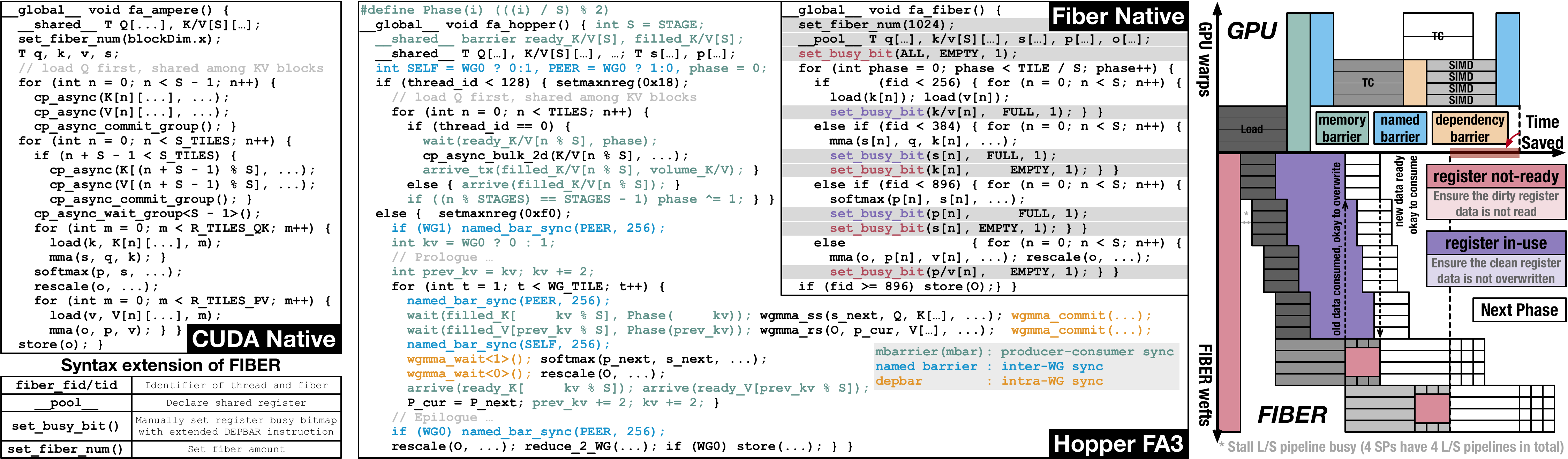}
    \vspace{-0.2cm}
    \caption{(left) FlashAttention-2 of \emph{CUDA-native mode}, (middle) FlashAttention-3 on Hopper, (middle-top) Attention in \emph{\proj{}-native mode}. Grey shaded lines mark \proj{} extensions. (right) Execution timeline diagram of Hopper FA3 and \proj{}.}
    \vspace{-0.2cm}
    \label{fig-program}
\end{figure*}

\section{\proj{} Software Stack}
This section introduces the \proj{} software stack, including the programming model illustrated through a FlashAttention example and the necessary compiler support.

\subsection{\proj{} Programming Model}
\label{sec:programmingmodel}
\proj{} programs are also organized into kernels, but adopt a \emph{thread-fiber} model, as illustrated in \Fig{fig:executionmodel}.
Each kernel launches a set of \proj{}-threads onto an SM, where each thread manages on-chip resources and spawns fibers that share the same physical register file and are responsible solely for computation.
An SM can host multiple \proj{}-threads with exclusive resources, analogous to SIMT-blocks.
While \proj{}-threads still own private registers, fibers—the parallel execution instances-are decoupled from register ownership.
Fibers are scheduled in 32-lane wefts with independent thread scheduling (ITS) semantics~\cite{CUDAProgramingGuide}, analogous to SIMT warps.

\proj{} first provides a \emph{CUDA-native} mode that enables a seamless transition from SIMT.
As depicted in \Fig{fig-program}(left), most SIMT CUDA kernel code remains unchanged.
Registers are treated as privately owned by fibers through the standard declaration, and same number of fibers as original CUDA thread are spawned with a \texttt{set\_fiber\_num()}. 

\Fig{fig-program}(middle-top) depicts the full power of \proj{} in \emph{\proj{}-native} mode.
It gives programmers a shared register view: tensors declared with \texttt{\_\_pool\_\_} are shared by all wefts, which can operate on them cooperatively to enable register-level dataflow. For example, a computed attention score \texttt{s} is reused directly by softmax wefts through registers, without shared memory round trip.
This also eliminates the tradeoff between register allocation and parallelism; in contrast, Hopper (\Fig{fig-program}(middle)) relies on \texttt{setmaxnreg} to manually partition registers among warp groups.
Programmers are also provided with a dynamic parallelism semantics: they can spawn a large number of fibers and assign them to heterogeneous tasks, forming (weft) specialized pipelines, and spatially unrolls the task into a dataflow execution paradigm.

Furthermore, \proj{} orchestrates weft-level fine-grained overlap and dataflow without multi-level synchronization. Hopper kernels rely on a mix of primitives: \textcolor[RGB]{79,129,119}{\textbf{memory barriers}} for producer-consumer synchronization, \textcolor[RGB]{25,128,192}{\textbf{named barriers}} for inter-warp-group coordination, and \textcolor[RGB]{216,130,32}{\textbf{dependency barriers}} for intra-warp-group coordination.
In \proj{}, different compute phases share the same register view and data flows directly at the register level, assisted by the fine-grained dependency tracking mechanism.
\Fig{fig-program} (right) illustrates the timeline of attention on Hopper (FA3) and \proj{}.
Two types of \texttt{set\_busy\_bit()} statements in \proj{} (lowered to extended \texttt{DEPBAR} instructions) manipulate different groups of the register busy bitmap to coordinate fine-grained dataflow, corresponding to \textcolor[RGB]{198,88,108}{\textbf{not-ready}} and \textcolor[RGB]{125,105,179}{\textbf{in-use}} in the timeline.

Setting the \textcolor[RGB]{198,88,108}{\textbf{\emph{empty}}} group prevents consumers from reading unready data, as there is no inter-wefts ordering guarantee. Initially, all entries are manually set, indicating no data is ready (in practice, an ``ALL''-scope setting is encoded into the reserved bits of \texttt{SET\_FIBER\_NUM} instruction, which serves as a synchronization point). Entries are cleared automatically by hardware once data is produced (written back). After the data is consumed, entries are manually set again to prevent the next phase from reading stale data.
Setting the \textcolor[RGB]{125,105,179}{\textbf{\emph{full}}} group prevents cross-phase overwriting. After a producer issues instructions for a pipeline stage, it sets the destination entry in this group, indicating the data is in use and stalling the producer's next phase from overwriting the buffer. Once consumers finish reading and enter the execution stage, the entry is automatically cleared by hardware, allowing new data to be written.
All orchestrations occur at weft granularity through shared registers, with minimal cross-weft synchronization overhead.

\subsection{Compiler Support}
\label{subsec:compiler_support}
In \emph{CUDA-native} mode, we apply a source-to-source transformation to the SASS emitted by \texttt{nvcc} and insert a \texttt{SET\_FIBER\_NUM} instruction into the kernel. 
A fixed offset of $\mathit{ID}_{\mathrm{subpartition}} \times 512$ is maintained at the decode stage and automatically added to each original operand identifier, emulating a subpartition-local private register ownership. 
Applying the \texttt{\_\_pool\_\_} qualifier activates \emph{\proj{}-native} mode, which exposes a shared register view. Here the compiler maintains a free list over 2048 vector registers, allocating on first assignment and releasing at end of lifetime via standard liveness analysis~\cite{LifeSpan}. 
It also performs operand placement for MMAs to minimize shared-register bank conflicts and imbalance, following predefined rules derived from the regular access patterns of GEMM. 
When a write targets a remote-subpartition register, the compiler rewrites it: the destination becomes a local register, and dependency analysis redirects consumer operands through cross-subpartition loads.

 \begin{figure*}[t]
    \centering
    \includegraphics[width=0.99\linewidth]{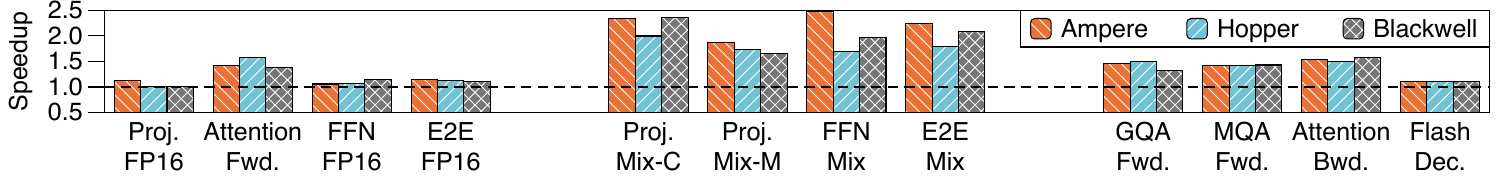}
    \vspace{-0.2cm}
    \caption{Workload speedups of \proj{}. Dotted line marks the Ampere, Hopper, and Blackwell baselines, respectively.}
    \vspace{-0.2cm}
    \label{fig:eval-all}
\end{figure*}

\begin{table}[b]
    \caption{Simulator configurations.}
    \vspace{-0.2cm}
    \label{tab:evalconf}
    \centering
    \footnotesize
    \begin{tabular}{cc}
    \toprule
    \textbf{Parameter} & \textbf{Value/Configuration} \\
    \midrule
    Subpartitions (SP) & 4 / SM \\
    Register File, Scheduler & 64 KB / SP, 1 (GTO) / SP \\
    ALU Lat., Throughput & SIMD,4,1 $\vert$ Tensor,24,1/(4,2,1) $\vert$ SFU,30,1/8 \\
    Mem Lat., Bandwidth$^*$ & Shared,30,128B $\vert$ L2,400,5120B~\cite{Ampere}\\
    \bottomrule
    \end{tabular}\\
    \vspace{0.1cm}
    $^{*}$ Bytes per cycle; Shared is per-SM, L2 cache is across the GPU.
\end{table}

\section{Evaluation}
In this section, we first describe the evaluation setup. We then justify \proj{}'s hardware overhead and present speedups of diverse workloads. Finally, we conduct an ablation study of \proj{}'s speedups with performance counter results.

\subsection{Experimental Setup}

\textbf{Simulation.}{ }
We use GTSim~\cite{GTSim}, a tile-graph-based GPU simulator, to model both baselines and \proj{}. Since the performance of modern GPU kernels is governed more by dependencies across tile-level computation and data movement, rather than by instruction latency alone, GTSim abstracts execution as a dependency-driven graph of warp-aligned data tiles, and thereby accurately captures the highly overlapped execution dataflow of the latest GPUs, powered by warp specialization, multi-stage pipelines, TMA, warp-group asynchronous TCs, and memory-based TC operand supply. 
It models diverse LLM workloads across GPU generations with a mean absolute percentage error (MAPE) of 1.2--11.3\%. 

Better overlapping and efficient dataflow execution are exactly what \proj{} sets out to enable, so GTSim is well suited to modeling the performance of \proj{}. 
We first enable SM-wide pooled registers in the hardware model, allowing register tiles to be accessed by other warps' operator nodes. The scheduler now maintains a busy bitmap per vector register, stalls instructions on unresolved dependencies, and arbitrates register accesses; scheduling contexts no longer carry register ownership. 
Dynamic parallelism and fine-grained register dataflow are expressed in programs composed in \emph{\proj{}-native} mode, by unrolling compute phases across more specialized wefts nodes, mapping compile-time-deterministic dataflow that is independent of runtime information to the shared register file, and inserting the corresponding full/empty state update instructions. 
\Tbl{tab:evalconf} lists the configuration details; a throughput of 1/$x$ means one instruction completes every $x$ cycles.

\textbf{Baseline. }
We use \emph{CUDA-native} as the baseline for SIMT GPUs and report execution-cycle speedups of \emph{\proj{}-native} modes. Warp specialization, warp-group TC, and asynchronous pipelining are enabled for Hopper and Blackwell baselines, with 2$\times$ and 4$\times$ TC performance than Ampere. 

\textbf{RTL and Synthesis.}
We implement the new units in Verilog and synthesize them with Synopsys Design Compiler using a TSMC 16~nm standard-cell library at 2.5~GHz; all components meet timing at this frequency. Crossbar area and power are instead modeled with ORION 3.0~\cite{ORION3} to account for wiring and fan-in/out overhead. 
We report area, power, and latency scaled to a 7~nm node~\cite{DeepScaleTool} for comparison against the A100.

\begin{table}[h]
    \caption{Evaluated Workloads}
    \vspace{-0.2cm}
    \label{tab:workloads}
    \centering
    \footnotesize
    \begin{tabular}{cccc}
        \toprule
        \textbf{Phase} & \textbf{Operation(s)} & \specialcell{\textbf{Configuration}\\\textbf{(Prefill)}} & \specialcell{\textbf{Launch}\\\textbf{Parameters}} \\
        \midrule
        QKV Proj. & GEMM & ($\emph{L}_{\emph{context}}$, $\emph{d}_{\emph{head}}$, & $\emph{KV}\_\emph{Block}=$ 64 \\
        Attention & RoPE,MHA & $\emph{d}_{\emph{hidden}}$, $\emph{d}_{\emph{ffn}}$) = & $\emph{blockDim}=$ \\
        Out Proj. & GEMM & ($\emph{4096}$, $\emph{128}$,& 256 (Proj., FFN), \\
        FFN Block & +SiLU,MUL & $\emph{4096}$, $\emph{11008}$) & 256/384/512 (FA2/3/4) \\
        \midrule
        \multicolumn{2}{c}{\textbf{Optimizations}} & \multicolumn{2}{c}{\textbf{Newly introduced operation(s)}}\\
        \midrule
        \multicolumn{2}{c}{Model Compression} & \multicolumn{2}{c}{Mixed-Precision GEMM (W4A16)} \\
        \multicolumn{2}{c}{Efficient Attention} & \multicolumn{2}{c}{Group/Multiple Query Attention} \\
        \bottomrule
    \end{tabular}
\end{table}

\textbf{Workloads.}
We evaluate \proj{} on LLM workloads using configurations from the Llama model family~\cite{llama}.
\Tbl{tab:workloads} details the phases evaluated in the prefill stage of inference: projection (QKV and output), attention (MHA), and feed-forward network (FFN). Interleaved vector operations such as RoPE and SiLU are also modeled.
On Ampere, Hopper and Blackwell, the attention is FA2~\cite{FlashAttention2}, FA3~\cite{FlashAttention3} and FA4~\cite{FlashAttention4}, respectively. 
We further evaluate kernels from two categories of LLM optimizations.
The first is mixed-precision (W4A16) GEMM for model compression, including AWQ~\cite{AWQ} (dequantization via vector operations) and AQLM\#~\cite{AQLM} (dequantization via codebook looking-ups).
The second is efficient attention variants: MQA and GQA with eight heads per group~\cite{GQA}, which reduce KV-cache footprint through cross-head sharing.
We also evaluate decoding and training workloads using FlashDecoding~\cite{FlashDecoding} and FlashAttention-2/3/4 backward, respectively. Each with $L_{\text{context}} = 4096$ and $d_{\text{head}} = 128$.

We mirror the workloads' open-source implementations~\cite{FlashAttention2,FlashAttention3,FlashAttention4,Tilelang} and feed them into the simulator. 

\begin{table}[h]
    \caption{Area, Power and Latency Overhead}
    \vspace{-0.2cm}
    \label{tab:RTL}
    \centering
    \footnotesize
    \begin{tabular}{cccc}
        \toprule
        \textbf{Item} & \textbf{Area (um$^2$)} & \textbf{Power (mW)} & \textbf{Latency (ns)} \\
        \midrule
        Read Port Arbiter    & 76.8   & 0.386  & 0.34 \\
        Register Busy Bitmap & 4910.4 & 8.649  & 0.25 \\
        Selector             & 3.04   & 0.003  & 0.09 \\
        Crossbar             & 317.9  & 1.688  & 0.12 \\
        \midrule
        Total (Latency.max)                & 5308.1 & 10.75  & 0.34 \\
        \% of A100 (108 SMs)    & 0.069\% & 0.39\% & 0.34/0.71*\\
        \bottomrule
    \end{tabular}\\
    \vspace{0.1cm}
    $^{*}$ Tesla A100's core frequency is 1.41 GHz~\cite{Ampere}, thus 0.71 ns per cycle.
\end{table}

\subsection{Hardware Overhead Analysis}
\label{sec:hwoverhead}
\proj{} reuses most baseline components, including IF/ID/WB, the physical register organization, and the execution pipeline. 
To support the shared register abstraction, it adds a selector, crossbar, and arbiter; and a register busy bitmap to enable fine-grained register dataflow scheduling. 

\Tbl{tab:RTL} lists the area, power, and latency of new components. In total they add 0.0053~mm$^2$ of area and 10.75~mW of power per SM, under 0.1\% and 0.4\% overhead, respectively.
The register busy bitmap logically comprises two groups of 2048 bits. To reduce port-wiring overhead, we replicate it across four subpartitions, each with fewer ports, giving two groups of 2048$\times$4 bits; a write to any entry is broadcast to all replicas consistency. 
The crossbar uses a 32-lane organization aligned with the vector registers, for which ORION~3.0 estimates a small area and power footprint. Moreover, crossbars of comparable scale already appear in current GPUs for L2 and shared-memory interconnects~\cite{Volta,Ampere,Hopper}.

Because these new components are placed in parallel with existing pipeline stages (\Fig{fig-uarch} (b)), they introduce negligible timing overhead and can meet upto 2.5~GHz frequency target.

In case of a tight wiring and routing budget, \proj{} can be simplified: enabling a single extra address bit reduces cross-subpartition register sharing (and its crossbar) to sharing within a subpartition. \textbf{DP} is still guaranteed at four-weft granularity, since fibers and registers remain decoupled. Register dataflow is confined within a subpartition, enforced by the programming model, which exposes a shared-register view among wefts with the same $\emph{ID}_{\emph{weft}}~\emph{\%}~\emph{4}$, and by the compiler, which lowers cross-subpartitions access to shared memory. 

\subsection{Speedup Results and Analysis}
\Fig{fig:eval-all} reports \proj{} speedups of diverse workloads against Ampere, Hopper and Blackwell baselines. 

\textbf{Speedup of LLM Workloads. }
On the Ampere baseline, \proj{} achieves a \textbf{1.42$\times$} speedup on forward attention, while projection and FFN show marginal gains (1.1$\times$), as these phases are dominated by GEMM kernels whose throughput is constrained by TC capacity rather than register or scheduling inefficiencies. In typical W4A16 LLM serving scenario aligned with vLLM~\cite{PagedAttention}, \proj{} delivers a \textbf{2.25$\times$} end-to-end speedup (\textbf{1.15$\times$} for FP16). 
On Hopper and Blackwell, similar trends hold, with \proj{} achieving \textbf{1.57$\times$} and \textbf{1.4$\times$} speedups in forward attention, and both correspond to $>$95\% TC utilization. Projection sees negligible improvement, as memory supplied operands already saturate the pipeline for pure GEMM. FFN gains are slightly higher (1.08$\times$, 1.14$\times$) due to interleaved SIMD and SFU operations beyond GEMM. Mixed-precision end-to-end improvements reach \textbf{1.8$\times$} and \textbf{2.1$\times$} (\textbf{1.13$\times$} and \textbf{1.11$\times$} for FP16). Notably, these speedups are achieved without increasing TC performance, but by closing the gap between achieved utilization and theoretical peak.




\textbf{Optimized LLM Workloads Speedup. }
Mix-C/M denote mixed-precision W4A16 GEMM variants, where ``C'' corresponds to \ul{C}ompute-intensive (AWQ) and ``M'' corresponds to \ul{M}emory-intensive (AQLM) dequantization.  
Both benefit substantially from \proj{}, achieving 2.33$\times$ and 1.88$\times$ speedups against Ampere. 
These speedups persist on Hopper and Blackwell.
The mixed-precision FFN benefits more than FP16 version due to more interleaved element-wise operations, achieving 2.49$\times$, 1.7$\times$, and 1.97$\times$ speedups. 
For attention variants, GQA achieves 1.45$\times$, 1.5$\times$, and 1.31$\times$ speedups, while MQA achieves 1.47$\times$, 1.51$\times$, and 1.46$\times$ speedups. 

\textbf{Decoding and Training Workloads Speedup.}{ }
For training workload, FlashAttention backward is still a TC-heavy workload: beyond having more GEMM than the forward pass, it keeps more live tensors on-chip and requires additional operations such as operand transposes and score recomputation.
Across Ampere, Hopper, and Blackwell, \proj{} delivers 1.54$\times$, 1.50$\times$, and 1.59$\times$ speedups over the FA2, FA3, and FA4.
These results correspond to nearly 95\% TC utilization, demonstrating that \proj{} scales well on training workloads.

For decoding workload, \proj{} delivers a 1.12$\times$ speedup.
This workload lies outside \proj{}'s primary target: FlashDecoding is dominated by GEMV and Softmax, which are highly memory-bound and do not use the TC even in the multi-batch setting (with distinct KV caches).
Although the baseline is already highly parallelised and \proj{} is designed to scale more powerful TCs, it still enables register-level dataflow for intermediate data that would otherwise spill to shared memory, along with more suitable per-phase parallelism.

\begin{figure}[t]
  \centering
  \begin{minipage}[b]{0.27\textwidth}
    \centering
    \includegraphics[width=\linewidth]{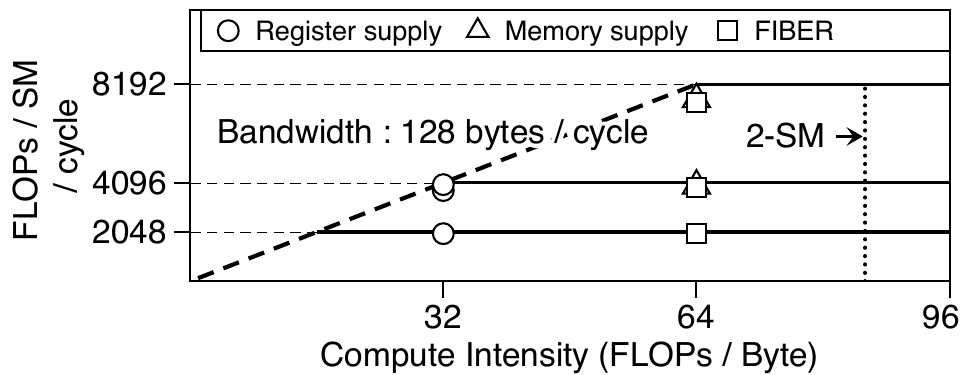}
    \caption{Roofline analysis.}
    \label{fig:rfline}
  \end{minipage}
  \hfill
  \begin{minipage}[b]{0.21\textwidth}
    \centering
    \includegraphics[width=\linewidth]{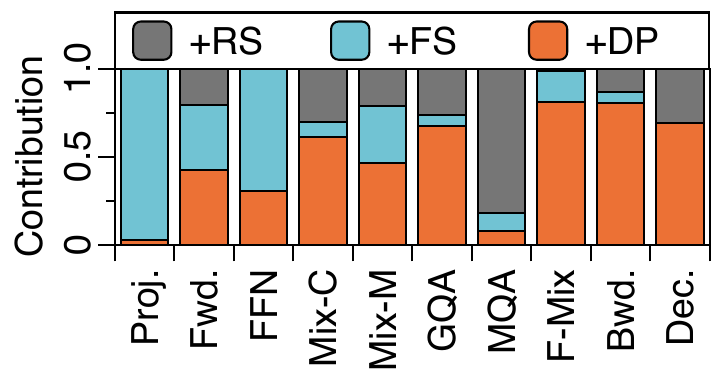}
    \caption{Speedup ablation.}
    \label{fig:ablation}
  \end{minipage}
\end{figure}

\textbf{Justifying Shared Register Supply for TCs.}{ }
\proj{}'s shared-register view can also provide redundancy-free operand supply for TCs. Here we justify this choice with GEMM.
On Ampere, whose TC delivers 2048~FLOPs/SM/cycle, the baseline register operand supply, which is not redundancy-free, already operates near saturation, as shown in \Fig{fig:rfline}.
When the TC scales to 4096~FLOPs/SM/cycle, private register supply reaches the turning point between the compute- and memory-bound regimes, and in practice its control overhead prevents it from fully exploiting the throughput.
When the TC scales further to 8192~FLOPs/SM/cycle, private register supply becomes clearly memory-bound.
To keep pace with evolving TCs, Hopper and Blackwell adopt shared/tensor-memory operand supply, which eliminates intra-SM data duplication and doubles compute intensity; Blackwell further introduces cross-SM data sharing~\cite{PTX90}.
Shared registers remove the same intra-SM duplication, likewise doubling compute intensity and keeping pace with evolving TCs. Although conceptually similar to shared-memory supply, register sharing additionally enables direct data reuse at the power-efficient register level, and decoupled register ownership as the key enabler of \textbf{DP} leads naturally to this shared-register supply alternative. 
Looking ahead, Blackwell's tensor memory can be viewed at the hardware level as an extension of the register file with additional banks, and \proj{}'s unified shared-register view can reduce programming burden and data-movement overhead while retaining the potential to scale further.

\subsection{Ablation Study}
\begin{figure}[t]
    \centering
    \includegraphics[width=0.725\linewidth]{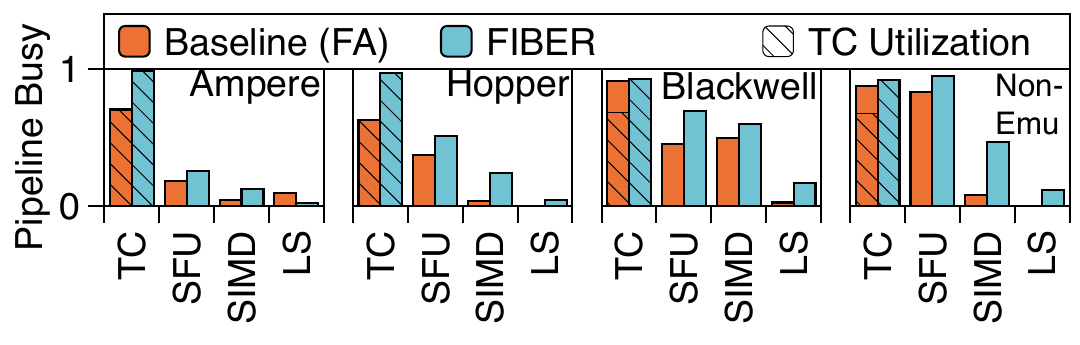}
    \includegraphics[width=0.25\linewidth]{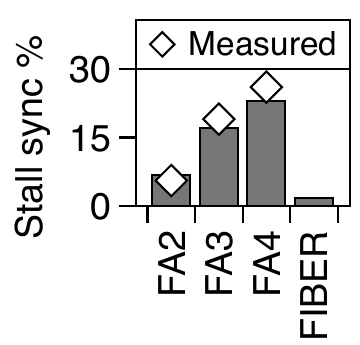}
    \caption{Performance counters of FA-Fwd.: (left) pipeline occupancy; (right) stall-sync \% of total CPI. Non-Emu: FA4 without \textbf{exp} numerical emulation for pipeline load balancing.}
    \label{fig:perfcnt}
  \end{figure}

\Fig{fig:ablation} presents an ablation study of the speedup contributions from \proj{}'s key features relative to the Hopper baseline.
\Fig{fig:perfcnt} reports the supporting performance counters. 

\textcolor[RGB]{186,74,17}{\textbf{$+$DP}} relaxes the fixed weft-count constraint while preserving Hopper's warp-group graunlarity, phase scheduled MMAs and shared memory based intermediate data reuse. It benefits nearly all evaluated workloads by up to 80\% except pure GEMM-based projection, because modern workloads typically interleave GEMMs with a diverse of other operations. \textbf{DP} is the most natural default capability enabled by decoupled register execution model of \proj{}. Even under an extremely constrained hardware budget, where the absence of a crossbar precludes SM-wide register sharing and the lack of busy bitmaps prevents fine-grained scheduling, \textbf{DP} remains effective, requiring only one bit extension to the register-address.

\textcolor[RGB]{54,155,175}{\textbf{$+$FS}} introduces a fine-grained dependency-tracking mechanism and scheduling on top of \textbf{DP}. Half of the workloads gain a clear speedup from \textbf{FS}. Although projection shows a small speedup overall, it benefits the most from \textbf{FS}, because aside from the MMA compute its baseline consists almost entirely of synchronization, which is exactly what \textbf{FS} eliminates. Other GEMM-bound workloads, such as FFN (69\%) and look-up-based mixed-precision GEMM (33\%), also benefit noticeably from \textbf{FS}. As more non-GEMM operations are interleaved, \textbf{FS}'s contribution is surpassed by \textbf{DP}. \Fig{fig:perfcnt} (right) further confirms that \textbf{FS} effectively reduces stall overhead.

\textcolor[RGB]{82,82,82}{\textbf{$+$RS}} additionally enables register-level dataflow via intra-SM register sharing: it supports shared-register-based TC operand supply and lets operations exchange and reuse intermediate data in registers. This yields a marked speedup across the evaluated workloads, and especially for those with heavy interleaved operations, most notably 80\% for MQA, whose MMA load is very light compared with its other operations.

\Fig{fig:perfcnt} (left) demonstrates that \proj{} improves both occupancy and achieved utilization of compute pipelines. 
\section{Related Work}
We summarize related work from following aspects.

\textbf{Execution Flow Optimization. }
A large body of work~\cite{FlashAttention3,FlashAttention4,GPUDirectAsync,AgileGPUSSD} improves GPU execution flow to better sustain TC utilization through asynchronous execution. Warp specialization~\cite{WASP,Singe,LiquidGeMM,Tawa,Twill} partitions heterogeneous tasks across warps to overlap them. Persistent kernels optimize latency-sensitive LLM serving~\cite{PTB,FractionGPU,MPK}. SM-wide dataflow, including cross-SM overlap and interconnect-enabled fusion, pushes the efficiency frontier further~\cite{Kitsune,ClusterFusion,FlashFuser,FlatAttn}. These techniques demonstrate the benefits of decoupling compute and memory to improve overlap and scheduling flexibility. \proj{} follows the same philosophy of decoupled, overlapped execution, but extends it to a finer granularity with a simple programming abstraction.

\textbf{Register Reorganization. }
Prior work explores register sharing, remapping, and reuse to exploit cross-warp operand reuse~\cite{RegMutex,SHREG,Duplo,INTERPRET}. These techniques share \proj{}'s goal of improving register efficiency but remain localized, whereas \proj{} provides a unified, SM-wide shared-register abstraction integrated into the execution model.

\textbf{Tensor Core Integration. }
Virgo~\cite{Virgo} and NVIDIA Blackwell~\cite{Blackwell} integrate TCs with SRAM at cluster granularity. RISC-V GPGPUs~\cite{RISCVGPU} and Intel Gaudi~\cite{Gaudi} use large cooperative tensor engines to improve operand reuse. These designs reflect a broader shift toward relaxing the coupling between parallel execution instances and register resources. \proj{} advances this trend more fundamentally by introducing lightweight fibers without private register ownership.

\textbf{Parallel Execution Model. }
To handle growing data volumes and increasingly irregular parallelism, dataflow-oriented architectures have regained attention. RipTide~\cite{RipTide}, Tyr~\cite{Tyr}, WaveScalar~\cite{Wavescalar}, and TTDA~\cite{TTDA} employ dataflow graphs and tag-based execution to exploit fine-grained parallelism and data reuse. Drawing on these abstractions while retaining compatibility with conventional GPUs, \proj{} extends SIMT by decoupling register ownership from parallel execution. This enables cooperative data reuse and dynamic parallelism while preserving a familiar GPU programming model.

\section{Conclusion}
We present \proj{}, an architecture built around a lightweight parallel execution abstraction for orchestrating tensor computation efficiently on GPUs with increasingly powerful tensor cores. 
By decoupling private register ownership from the parallel execution instance (fiber), \proj{} lets all fibers cooperatively access a shared register file, supporting dynamic parallelism scaling, fine-grained dataflow scheduling, and register-level operand reuse and dataflow orchestrating. 
\proj{} combines ISA extensions, minimal microarchitectural enhancements, and a SIMT-compatible programming model, achieving \textbf{1.8$\times$--2.3$\times$} end-to-end speedups across GPU generations in representative mixed-precision LLM serving scenarios, with kernel-level speedups of up to \textbf{2.49$\times$}.



\bibliographystyle{IEEEtranS}
\bibliography{refs}

\end{document}